\documentclass{aa}
\usepackage{graphicx,natbib}

\newcommand{\kmspc}{\mbox{\,km\, s${}^{-1}$\, pc${}^{-1}$}}
\newcommand{\msun}{\mbox{\,M${}_{\sun}$}}
\newcommand{\kms}{\mbox{\,km\, s${}^{-1}$}}
\newcommand{\kpc}{\mbox{\,kpc}}
\newcommand{\pc}{\mbox{\,pc}} 
\newcommand{\bfr}{\mbox{\boldmath $r$}}
\newcommand{\vel}{\mbox{\boldmath $\dot{r}$}}
\newcommand{\acc}{\mbox{\boldmath $\ddot{r}$}}
\newcommand{\del}{\mbox{\boldmath $\nabla$}}
\newcommand{\cross}{\mbox{\boldmath $\times$}}

\begin{document}

\title{Dynamical Modeling of the Stellar Nucleus of M31}

\author{Niranjan Sambhus \and S. Sridhar}
\offprints{Niranjan Sambhus, \email{nbs@iucaa.ernet.in}}
\institute{Inter--University Centre for Astronomy and Astrophysics 
\\
 Ganeshkhind, Pune 411 007, India}

\date{\today}
\authorrunning{Sambhus \& Sridhar}
\titlerunning{Modeling the nucleus of M31}

\abstract{
We present stellar dynamical models of the lopsided, 
double--peaked 
nucleus of M31, derived from {\sl Hubble Space Telescope (HST)}
photometry. A Schwarzschild--type method, in conjunction with 
Richardson--Lucy deconvolution, was employed to construct
steadily rotating, hot, stellar disks. The stars orbit a massive 
dark
object, on prograde and retrograde quasi--periodic loop orbits.
Our results support Tremaine's eccentric disk model, extended to
include a more massive disk, non zero pattern speed ($\Omega$), and
different viewing angle. Most of the disk mass populated prograde 
orbits, with $\simeq 3.4\%$ on retrograde orbits. The best fits to
photometric and kinematic maps were disks with $\Omega\simeq
16\kmspc\,$. We  speculate on the origins of the lopsidedness,
invoking recent work on the linear overstability of nearly 
Keplerian
disks, that possess even a small amount of a counter--rotating
component. Accretion of material---no more massive than a globular
cluster---onto a preexisting stellar disk, will account for the 
mass
in our retrograde orbits, and could have stimulated the 
lopsidedness
seen in the nucleus of M31.   

}
\maketitle


\section{INTRODUCTION}

The nuclei of normal galaxies are thought to harbor massive dark
objects (MDOs), which could be supermassive black holes. These
central regions often possess dense agglomerations of stars, whose 
structural and kinematical properties appear to be correlated with 
global galaxy properties \citep{geb96,fm00,geb00}. The imprint of 
galaxy formation is surely recorded in the nature of stellar 
orbits.
No more unusual examples are, perhaps, known than the nuclei of 
the 
galaxies, NGC4486B (in the Virgo cluster) and M31 (our nearest 
large
neighbor). The proximity of M31  has enabled detailed photometric
and kinematic observations of its nucleus, beginning with the
detection of its asymmetrical shape by Stratoscope~II 
\citep{lig74},
and its resolution into a double--peaked structure by the {\sl HST}
images of \citet{lau93}. The central peak (P2) lies close  to the
presumed location of the MDO, located in a small region of 
UV--bright
stars  \citep{kin95,lau98,kb99}. \citet{tre95} proposed that the 
off--centered peak (P1) marks the region in a disk of stars, where 
lie 
the apoapses of many eccentric orbits. This lopsided structure is
expected to rotate steadily with some pattern speed, and remain 
locked in place by the self--gravity of all the stars.
We construct numerical  stellar dynamical models, wherein the disk
potential is derived directly---after bulge subtraction---from the
{\sl HST} photometry of \citet{lau98}. Model construction and 
comparisons with data make many demands on computational 
resources. 
Hence it was not practical to explore the effects of varying 
values of many of the parameters concerning the bulge and disk;
we take many of these values from \citet{kb99}. However, we do 
explore 
the effect of varying the pattern speed. We state our assumptions 
and
give an outline of our method below.

We assumed that the bulge--subtracted light emanated from a 
steadily
rotating, inclined, razor--thin, flat, disk of stars, in orbit 
about
the MDO. The stars compose a collisionless, self--gravitating
system. Hence the orbits of individual stars are governed by the
combined gravitational attractions of the MDO, and the smooth,
self--gravitational potential of all the stars. A bulge--disk
decomposition of the V--band image of \citet{lau98} yielded the
disk surface density, from which the smooth disk potential was
computed. For some chosen value of the pattern speed ($\Omega$),
orbits of test stars were integrated numerically in the rotating
frame. A selection of prograde and retrograde (quasi--periodic) 
loop
orbits of various sizes composed an orbit library. The orbits were
populated with ``stars'' ($\sim 237,000$ in all), spaced uniformly 
in
time, and the disk light (in a central region) partitioned into 
many
cells, with more cells than orbits. Determination of orbit masses,
from the known luminosities of the cells, required solving an
overdetermined problem, involving  positive quantities. This was
achieved through $\sim 5000$ iterations of a  RL
algorithm \citep{ric72,luc74}. The entire procedure was repeated 
for
several values of $\Omega$. Comparisons with the kinematic maps of
\citet{bac01} followed. Central line--of--sight velocity
distributions were calculated to emphasize the regions in velocity 
space, where retrograde orbits contribute. Following
\citet{kb99}, we assumed a distance to M31 of $770\kpc$ (on the 
sky,
$1\arcsec$ corresponds to $\simeq 3.73\pc$), mass of the MDO,
$M\,=\,3.3 \times  10^7\msun\,$, and mass--to--light ratio of the 
(bulge--subtracted) nuclear disk equal to $\Upsilon_V = 5.73\,$.

\section{MODEL CONSTRUCTION}

\subsection{Deprojection and Disk Potential}

\begin{figure}[t]
\resizebox{\hsize}{!}{\includegraphics{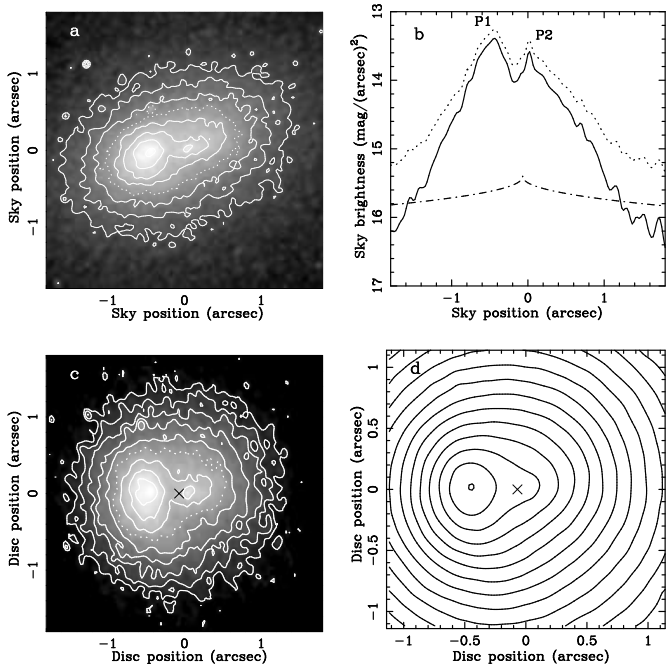}}
\caption{Derivation of the nuclear disk, and its (smooth)
gravitational potential. (a) Observed sky brightness distribution,
from \citet{lau98}; the dotted curve has magnitude $14.3$, and
successive isocontours differ by $0.25$ magnitudes. (b) Brightness
profiles along the P1--P2 line: observed (dotted curve), the 
S\'ersic
bulge (dashed--dotted curve), and the bulge--subtracted nuclear 
disk
(solid curve). (c) Brightness distribution of the disk, viewed
face--on;  the isocontours follow the same convention as in (a).
Isocontours of the disk--potential are displayed in (d); the units
are such that the deep minimum near P1 has depth equal to unity, 
and
successive isocontours mark increments of $0.05\,$. In (c) and (d),
``X'' marks  the location of the center of mass.}
\label{fig1}
\end{figure}

Fig.~1a shows the nucleus of M31, plotted from the  V--band, {\sl
HST} observations of \citet{lau98}. The UV cluster and the MDO are 
at
the origin. P2 is near the MDO, with sky coordinates ($0 \farcs 
023,\, 
0\arcsec$), and P1 is located at ($-0 \farcs 48,\,0\arcsec$). 
The bulge was assumed to be spherical, with a S\'ersic
brightness profile \citep{ser68,kb99}---see Fig.~1b. The center 
of
mass (COM) of the  bulge, disk, and MDO was required to coincide 
with
the bulge center; this common location was determined, by an
iterative method, at ($-0 \farcs 0684, 0\arcsec$), in agreement 
with
\citet{kb99}. With one notable exception, \citep{bac01}, all
investigations have assumed that the nuclear disk is coplanar with 
the much larger galactic disk of M31. We obtained very poor results
with this assumption. Therefore, we resolved to determine
the inclination and orientation of the nuclear disk, based on the
photometry, similar to \citet{bac01}. The disk light covered an
approximately elliptical region, with a ragged edge. The plane in
which the best--fit ellipse (to the edge)  deprojected to a circle
was defined to be the disk plane; its inclination ($i$), and PA of
the line of nodes, were $\simeq 51.54{\degr}$, and $\simeq
62.66{\degr}$, respectively. The face--on view of the disk, shown 
in
Fig.~1c, has mass $\simeq 2.15\times 10^7\msun\,$. 

To minimize edge--effects, the self--gravitational potential
was evaluated in the disk plane, at $10^4$ grid points within a
central square, of side equal to $2 \farcs 28$. However, the
Newtonian $|{\bf r} - {\bf r'}|^{-1}\,$ contributions from the
entire disk of Fig.~1c, which has diameter $\simeq 3 \farcs 6$, 
was used. The grid values were fit to a $20$--th order polynomial 
function of the Cartesian coordinates, $\Phi_d(\bfr)$, a contour 
plot of  
which is displayed in Fig.~1d. The polynomial form smoothed the
potential, facilitated coding of the integrator, and checking of  
the
integrated orbits in the nearly Keplerian limit \citep{st99}. 
Figs.~1c 
and 1d can be imagined as either snapshots of a rotating  
configuration,
or as steady images in a frame rotating with some angular speed 
$\Omega$,
about an axis normal to  the disk plane, and passing through the 
COM.  
The forces on a test star include the gravitational attractions of 
the
MDO and disk, as well as centrifugal and Coriolis forces. The 
contribution of the bulge was ignored, because it is so much 
smaller than
the other forces. 

\subsection{The Orbit Library}

\begin{figure}[t]
\resizebox{\hsize}{!}{\includegraphics{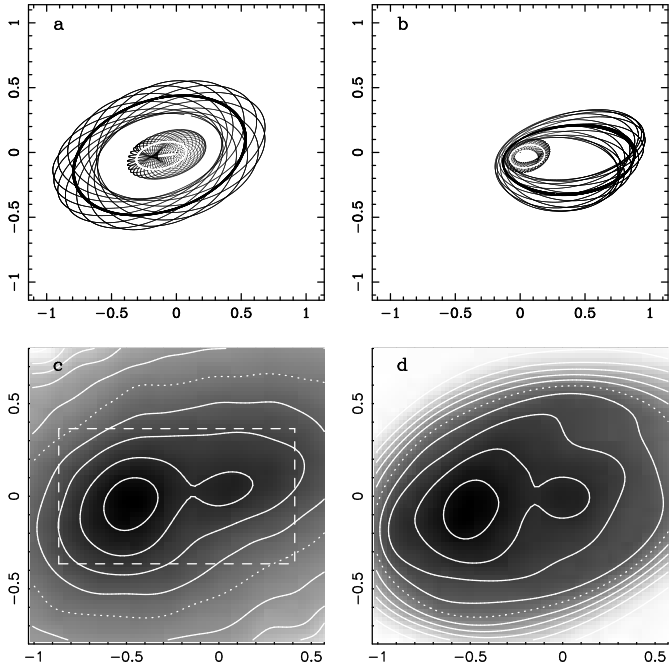}}
\caption{Orbits in the rotating frame and photometric fits.
The axes in all panels are sky positions. (a) and (b) show prograde
and retrograde loop orbits, respectively, as seen on the sky, for
$\Omega = 16 \kmspc$; the parent (resonant) orbits are overdrawn as
the solid curves. The photometry in (c) is from \citep{lau98},
smoothed with a Gaussian beam of FWHM~$=0 \farcs 17\,$. The
(bulge--subtracted) light in the region enclosed by the dashed box
was employed in our Schwarzschild--type iterative method. (d) is 
our
model disk, including the bulge. The dotted lines in both (c) and 
(d)
have magnitude equal to $14.5$, and successive isocountours differ 
by
$0.25$ magnitudes. The brightness is displayed in the ``negative''
mode, to better emphasize the distribution.}
\label{fig2}
\end{figure}

Orbits were computed in the rotating frame by numerically 
integrating
the equations of motion,
\begin{equation}
\acc \;=\; -\frac{GM}{r} \;-\; \del\Phi_d 
\;-\; \Omega^2\bfr \;-\;
2\Omega\,\left(\hat{z}\,\cross\,\vel\right)\,,  \label{eom}
\end{equation} 
\noindent using a $4$th--order, adaptive step size,
Runge--Kutta scheme. The global structure of orbits was explored by
studying Poincar\'e surfaces of section. The principal families of
orbits were lenses and loops. Lens orbits  change the sign of 
their orbital
angular momentum \citep{st97,st99}. Stars on such orbits will
collide with the MDO, in the time it takes an orbit to precess.
These time scales do not exceed a million years, even for 
quite large orbits; dwarf, as well as giant stars on lens 
orbits will be lost to the MDO (if not tidally disrupted
before). Hence lens orbits were excluded from our modeling. Other
orbits that were also omitted included chaotic orbits, and those
parented by higher order resonances. The loops orbits were of two
kinds: prograde and retrograde, some of which are shown in Figs.~2a
and 2b; these were the only orbits included in our orbit library. 
The
kinematic model of \citet{tre95}, the Kepler--averaged dynamics of
\citet{st99}, and studies of slow, linear modes by \citet{tre01}, 
all
suggest the use of the prograde loops as the back bones of the 
orbit
library. The necessity of including retrograde loops is less 
obvious,
and was stimulated by the investigations of \citet{tou01}. It 
turned
out that the retrograde loops significantly improved fits near P2.

For each value of energy, loops of two or three different
``thicknesses'' (i.e. deviations from the parent loop) were 
computed.
Each orbit was sampled, and populated with ``stars'', spaced apart
uniformly in time. All stars in an orbit are accorded  the same
(unknown) mass; this is not a restriction, because in  a
collisionless system, the relevant physical quantities are the mass
per orbit. The numbers of ``stars'' in an orbit was chosen
proportional to the inverse square of the energy (approximately, 
square of the ``semi--major axis'') of the parent loop; thus $25$
stars sufficed for an orbit with $a=0 \farcs 02$, whereas an orbit
with $a=0 \farcs 6$  was sampled by more than $10,000$ stars.
Altogether, the positions and velocities of $\sim  237,000$ stars,
populating $50$ prograde orbits and $20$ retrograde orbits, were
recorded. 

\subsection{Richardson--Lucy Deconvolution}

Orbit masses were determined by iteratively imposing on the model,
consistency with the bulge--subtracted sky brightness of a region
covered by the orbits; the dashed box of Fig.~2c encloses this
region. The box was divided into $112$ equal square cells, each of
side $0 \farcs 09\,$;  each cell was small enough to give good
resolution, and large enough (16 pixels) to keep pixel noise levels
low. The ``observed'' mass per  cell, $\mu_j$ (for $j=1\ldots 
112$),
was obtained from the observed light, by multiplication with
$\Upsilon_V\,$; these numbers composed our basic data. We defined
$m_i$ (for $i=1\ldots 70$) as the mass of orbit $i$, that also lies
within the box; the total mass in the orbit exceeds $m_i\,$. We 
normalized $\sum_{i =1\ldots 70} m_i = \sum_{j=1\ldots 112}\mu_j =
1\,$, to unit mass in the box.  A linear relationship, $\mu_j \,=\,
\sum_{i=1\ldots 70} K(j|i)\,m_i$, exists between the ``observed
masses'' $\mu_j$, and the unknown masses $m_i$. The positive 
kernel,
$K(j|i)$, is known from the orbit library. It has the property,
$\sum_{j=1\ldots  112}K(j|i)=1$, for all $i=1\ldots 70$.  An 
initial
guess, $\{m_i^{\rm g}\}$, was iterated by the RL algorithm
\citep{ric72,luc74}. The problem being overdetermined, about $5000$
iterations ensured good convergence to some $\{m_i^{\rm f}\}$.
Velocities were then transformed to the inertial frame. Rescaling 
of 
$\{m_i^{\rm f}\}$ to physical values, and including the
portions of orbits outside the box, provided a numerical 
distribution
function. The entire process, beginning from the selection of an
orbit--library, was repeated for several values of $\Omega$, 
between
$5$ and $25\kmspc\,$. For any chosen value of $\Omega$, the final 
set of orbit masses,
$\{m_i^{\rm f}\}$, corresponds to a prediction for  the cell 
masses,
$\mu_j^{\rm f} \,=\,\sum_{i=1\ldots 70} K(j|i)\,m_i^{\rm f}$, which
should be compared with the data, $\{\mu_j\}\,$. For models with 
$\Omega \,=\, \{15, 16, 17\}\,$, the root--mean--squared deviation 
in mass per cell are, $\{0.26, 0.20, 0.28\}\,$; other values of 
$\Omega$ resulted in very poor models.

\section{COMPARISONS AND CONCLUSIONS}

\begin{figure}[t]
\resizebox{\hsize}{!}{\includegraphics{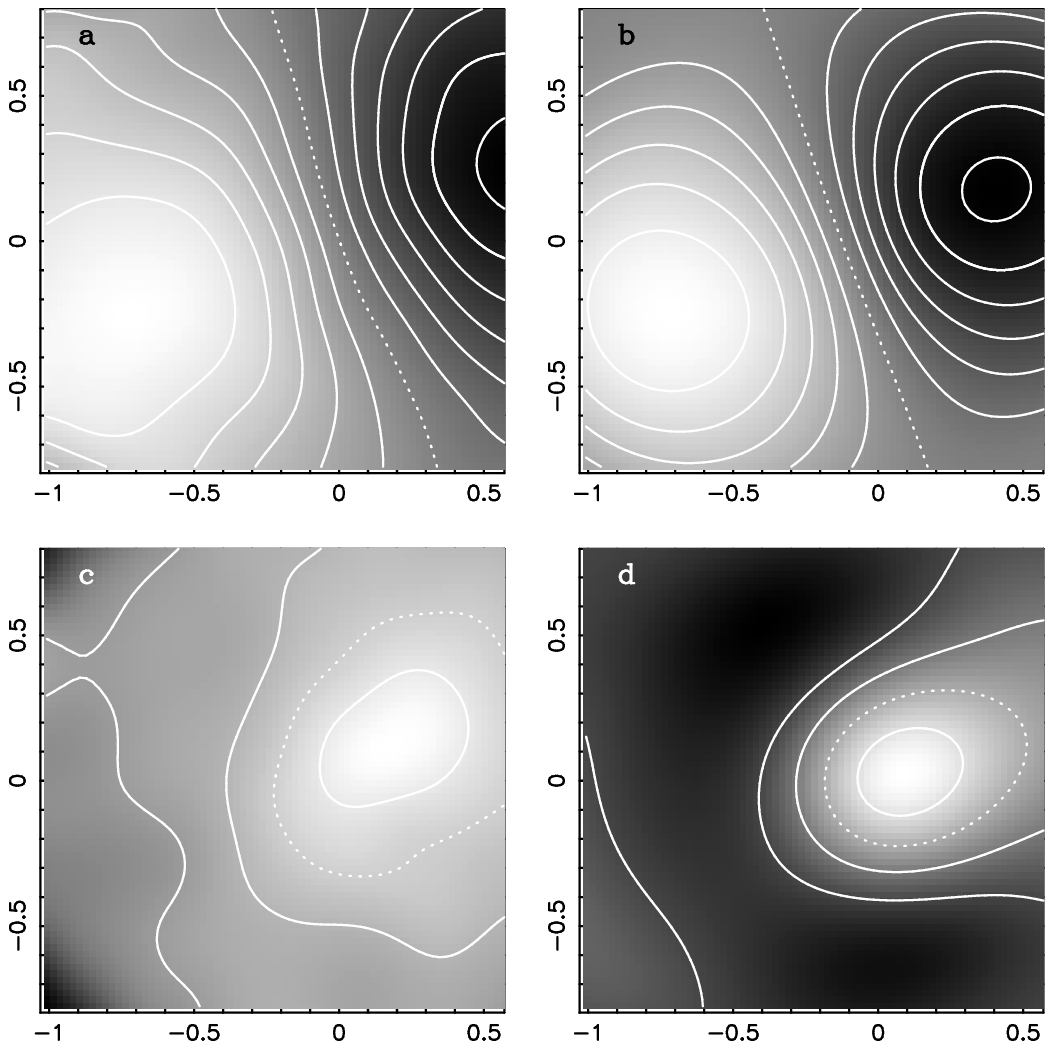}}
\caption{Comparison of the $\Omega = 16 \kmspc$ model with 
kinematic
maps. The axes in all panels are sky positions. (a) and (c) are 
maps of
mean line--of--sight velocity, and velocity dispersion,
respectively, taken from the ``M8'' data of \cite{bac01}; (b) and 
(d)
are predictions of our model,  including a constant bulge velocity
dispersion of $150\kms$, and smoothed with a Gaussian beam of 
FWHM~$=
0 \farcs 5$. In (a) and (b), the  dotted line is the 
zero--velocity curve,
and successive isocontours are in steps of $25\kms$; positive
(negative) velocities are in light (dark) shades. The dotted line 
in
(c) and (d) corresponds to velocity dispersion of $200\kms$,
successive isocontours are in steps of $25\kms$, and lighter shades
indicate higher values.}
\label{fig3}  
\end{figure}

We restored the S\'ersic bulge profile, for comparisons with the
photometry---see Figs.~2c and 2d, where the $\Omega = 16\kmspc\,$
model is compared with the photometry of \citet{lau98}. 
The locations of the peaks agree, although the model runs out of 
orbits near the edges. For kinematic comparisons, we further  
assumed that the velocity distribution of the bulge stars was
Gaussian,  with $\sigma_v=150\kms\,$. Fig.~3 compares
the model with the kinematic maps of \citet{bac01}. The need to
smooth with a beam of $FWHM = 0 \farcs 5$, rendered the absence of
the outer orbits more acute. However, the zero velocity curves, as 
well as the orientation of the line joining the maximum and minimum
velocities are in agreement (Figs.~3a and 3b); the dispersion maps 
are
reasonably compatible in the region of the peak near P2 (Figs.~3c 
and
3d). As noted earlier, the best fits obtained for models with
$\Omega = 15,\, 16,\,\mbox{and}\, 17\kmspc\,$. In Figs.~4a--4c, 
these
are compared  with the photometry of \citet{lau98}, and {\sl HST
STIS} kinematics from \citet{bac01}. Together, they should give 
some
idea of the deviations from observations. The pattern speed has 
been
variously estimated \citep{ss00,ss01,bac01} to lie between $3$ and
$25\kmspc\,$. Our present estimate, $\Omega\simeq 16\kmspc\,$, is
closest to \citet{ss01} who, however, prefer to view the disk at 
the
traditional inclination of $77{\degr}\,$.

We note here some limitations of our dynamical models. A basic 
assumption of our procedure was that the nuclear disk is 
razor--thin, and inclined at an angle of $(77^{\circ}- 
51.54^{\circ}) \simeq 25^{\circ}$, with respect to the plane of 
the larger galactic disk of M31. We also ignored the gravitational
force of the bulge stars on the nuclear disk, because the 
net effect of a spherically symmetric bulge would be to only modify
the precession rates by a small amount. However, it is known 
\citep{ken83,ken89} that the bulge of M31 is flattened, and this 
can be expected to modify the models in at least two ways.
If the flattened bulge were treated as a fixed, external 
potential, the node of the nuclear disk will precess. The more 
serious effect arises from the dynamical friction of the bulge,
acting on the  stars composing the nuclear disk. The torque
exerted by a flattened bulge, whose stars could have anisotropic 
distributions of velocities, could well decrease the inclination  
of the disk. However, we have not been able to estimate the 
response of the stellar disk, whose structure is so fundamentally
determined by eccentric orbits locked in resonance.

The assumption that the nuclear disk is razor--thin is, of course, 
unrealistic. \citet{tre95} estimates that two--body relaxation 
would thicken the disk significantly, within a Hubble time. Our 
choice of a razor--thin disk was made primarily for the recovery 
of a ``unique'' surface density distribution for the disk in its 
plane (Fig.~1c), from the observed surface photometry (Fig.~1a). 
This surface density was then used to calculate the disk 
self--gravity, possible orbit families for a range of pattern 
speeds, and then populating the orbit libraries appropriately 
using the RL algorithm. Consideration of a thick disk would have 
introduced an infinity of possible choices in the very first step 
of our procedure, and we wished to avoid it. It should, however, 
be stressed that an uninclined thick disk could well be compatible 
with observed photometry. This possibility should certainly be 
explored, perhaps by including kinematic data as additional 
constraints. A question that no one, presenting stellar dynamical 
models, can afford to ignore is whether the system is stable. 
There appears to be no better route to address this question, than 
N--body simulations. In this light we should regard the models 
presented in this paper as plausible guesses for further numerical 
explorations.

\begin{figure}[t]
\resizebox{\hsize}{!}{\includegraphics{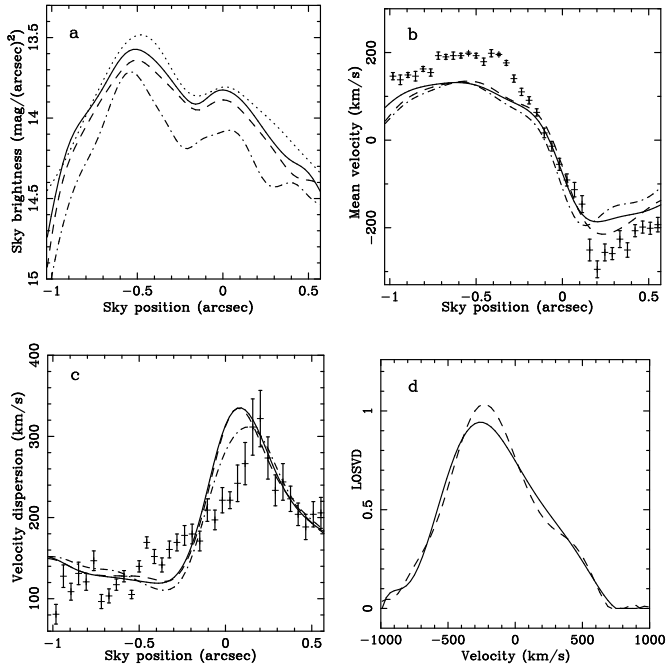}}
\caption{Further comparisons of models and observations. In (a), 
(b),
and (c),  the dashed--dot, solid, and dashed lines correspond to
predictions of our model disks (including the bulge), with $\Omega 
=
15,\, 16,\, \mbox{and}\, 17\kmspc\,$. The dotted line in (a) is
a cut along the P1--P2 line of the observed brightness, shown in
Fig.~3a. In (b) and (c), the plotted data points are HST
STIS  observations (from \citet{bac01}), of mean line--of--sight
velocity and velocity dispersion, respectively, taken with a slit 
of
width equal to $0 \farcs 1\,$, placed at PA of $39{\degr}\,$. The 
three
lines represent similar ``observations'' of our models.
In (d) we plot the LOSVD, observed with a Gaussian beam of FWHM~$=
0 \farcs 21$, for the $\Omega = 16\kmspc\,$ model
(including bulge), centered on the MDO. The dashed line was
computed after suppressing the retrograde orbits.}
\label{fig4}
\end{figure}

The ``eccentricity'' profiles of the loop orbits are given in 
Fig.~5a.
The prograde loops have a characteristic non monotonic profile,
whereas the retrograde loops have large eccentricites that 
increase monotonically with size, to the biggest orbits
employed in our models. We note that the eccentricity profile
of the prograde orbits is quite different from \citet{ss01}: in 
particular, there is no tendency for them to switch apoapses to the
anti--P1 side of the MDO. The profiles of the apoapse angles 
(Fig.~5b) show no evidence for spirality; prograde/retrograde loops
have  their apoapses on the P1/anti--P1 side of the MDO. The disk
mass is $1.4\times 10^7\msun$, with $3.4\%$ on retrograde orbits; 
the
central LOSVD in Fig.~4d indicate the positive and negative
velocities at which the latter contribute. We have tried models 
with
only prograde loops, but these gave consistently poor fits around 
P2. 

\begin{figure}[t]
\resizebox{\hsize}{!}{\includegraphics{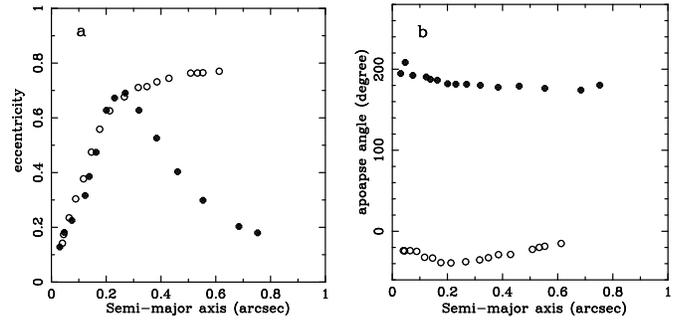}}
\caption{Distribution of orbital eccentricities and apoapse angles,
with orbit size. The ``semi--major axis'' is defined as the
mean of the maximum  and minimum radii ($r_{>}$ and $r_{<}$) of a
parent loop; $\mbox{eccentricity}\equiv (r_{>} - r_{<})/(r_{>} +
r_{<})\,$. In (a) and (b) the eccentricity and angle to apoapse, of
parent loops, are plotted for a selection of prograde (filled
circles) and retrograde (open circles) parent loop orbits. Prograde
(retrograde) parents have apoapses on the P1 side (anti--P1 side) 
of
the MDO.}
\label{fig5}
\end{figure}

Numerical simulations \citep{js01}, and analytical study 
\citep{tre01}
indicate that nearly Keplerian disks (without counter--rotating
streams) are neutrally stable to linear, $m\,=\,1$ perturbations. 
Hence it might appear unlikely that the lopsidedness could have 
grown spontaneously from an initially axisymmetric disk. Note,
however, that in \citet{bac01} there is reference to work, to be
reported in the future by Combes and Emsellem, on an  $m\,=\,1$
instability. \citet{bac01} also suggest that a lopsided mode  could
have been excited by the passage of a massive object, such as a 
giant molecular cloud, or a globular cluster. They also report
supportive simulations, where excited modes remained undamped 
for $7\times 10^7$ years, with almost constant pattern speed; this 
is certainly a plausible scenario.

Here we consider an alternative origin of the lopsidedness, based
on the presence of the retrograde loops in our models, and recent 
work 
by \citet{tou01} on a {\sl linear} instability, in a 
softened--gravity
version of Laplace--Lagrange theory of planetary motions. To the
extent that softened--gravity mimics the velocity dispersions of 
stars \citep{mil71,eri74}, this work suggests that even a few 
percent 
of mass in counter--rotating orbits could excite a linear $m\,=\,1$
overstability. In an axisymmetric nearly Keplerian disk, the 
apsides of
prograde/retrograde orbits have negative/positive precession rates.
A resonant response of the retrograde orbits (to a perturbation 
with
positive $\Omega$) appears to excite large eccentricities. This 
compensates for their small mass fraction, allowing them to act so
significantly, that the precession of apsides of the prograde 
loops 
is locked to that of the retrograde loops. We note that the large 
eccentricities of our retrograde loops (Fig.~5a), obtained directly
from orbit integrations, are suggestive of this possibility. We
speculate further that the overstability (as is common in other
contexts)  is arrested in growth by nonlinearity, and it settles 
into
a nonlinear, neutral mode. The steadily rotating nuclear disk of 
M31
might well be in such a phase. Material on retrograde orbits could
have been accreted by the infall of debris into the center of M31. 
One
possibility is suggested by our estimate of the mass in  retrograde
orbits in our models, $\sim 5\times 10^5\msun\,$. \citet{tre75} 
have
argued that dynamical friction would cause globular clusters to
spiral in toward galactic nuclei, and tidally disrupt. We could be
witnessing the lopsided signature of such an event.

\begin{acknowledgements}
We thank Drs.~S.~Faber and T.~Lauer for making available their 
{\sl HST} photometry, Dr.~E.~Emsellem for providing the {\tt OASIS}
kinematic maps, Dr.~J.~Touma for sharing results and thoughts, and
Drs.~R.~Nityananda and K.~Subramanian for advice and comments. 
We are also grateful to an anonymous referee for thoughtful
questions and comments. NS
thanks the Council of Scientific and Industrial Research, India, 
for
financial support through grant  2--21/95(II)/E.U.II.
\end{acknowledgements}

\end{document}